\documentclass[12pt]{iopart}
 \expandafter\let\csname equation*\endcsname\relax
 \expandafter\let\csname endequation*\endcsname\relax

\usepackage{amssymb, amsmath}
\usepackage{graphicx}

\DeclareMathOperator{\diag}{diag}
\DeclareMathOperator{\sgn}{sgn}

\DeclareMathOperator{\im}{Im}

\begin{document}

\title{On Random Matrix Averages Involving Half-Integer Powers of GOE Characteristic Polynomials}

\vskip 0.2cm
\author{Y.~V.~Fyodorov  and A. Nock}
\address{Queen Mary University of London, School of Mathematical Sciences, London E1 4NS, United Kingdom}

\begin{abstract}
Correlation functions involving products and ratios of half-integer powers of characteristic polynomials of random matrices from the Gaussian Orthogonal Ensemble (GOE)
frequently arise in applications of Random Matrix Theory (RMT) to physics of quantum chaotic systems, and beyond.
We provide an explicit evaluation of the large-$N$ limits of a few non-trivial objects of that sort within a variant of the supersymmetry formalism, and via a related but different method. As one of the applications we derive the distribution of an off-diagonal entry $K_{ab}$ of the resolvent (or Wigner $K$-matrix) of GOE matrices which, among other things, is of relevance for experiments  on chaotic wave scattering in electromagnetic resonators.
\end{abstract}

\maketitle

\section{Motivations, background and results}

\subsection{Introduction}

The goal of the present article is to attract attention to the problem of systematic evaluation of the large-$N$ asymptotics
of random matrix averages of the form
 \begin{equation}\label{1}
{\cal C}_{K,L}(\mu_{F1},\ldots,\mu_{FK};\mu_{B1}, \ldots,\mu_{BL})= \left\langle \frac{\det(\mu_{F1}-H) \dots \det(\mu_{FK}-H)}{\det^{1/2}(\mu_{B1}-H) \dots \det^{1/2}(\mu_{BL}-H)} \right\rangle_{GOE}
\end{equation}
where $\mu_{Fi}, \, i=1,\ldots,K$ and $\mu_{Bj}, \, j=1,\ldots,L$ are sets of complex parameters.
 The angular brackets here and henceforth denote the average over the ensemble of real-symmetric $N\times N$ matrices $H$ with Gaussian entries characterised by the probability density ${\cal P}(H)\propto \exp{-\frac{N}{4J^2}\mbox{\small Tr} H^2} $ and known as the Gaussian Orthogonal Ensemble (GOE). Note that the correlation functions involving products of
 square roots of the characteristic polynomials in the numerator can be always reduced to the above form by multiplying and dividing both the numerator and the denominator with the same corresponding factors.

 Although there are reasons to suspect that the correlation functions (\ref{1}) may have a nice mathematical structure even for finite $N$, perhaps not unlike those determinantal or Pfaffian structures discovered in  \cite{BH,FS03a,FS03b,BDS,BS06} for similar objects involving only integer powers (see also \cite{KG1,KG2} for an alternative derivation)  we were not yet able to reveal such structures beyond the simplest case $K=1,L=1$, see (\ref{8}) below. Instead we are mainly concentrating on the large-$N$ limit of a few simplest, yet nontrivial examples of the correlation function
of the type (\ref{1}). We start with considering correlation functions with  two square roots in the denominator, and with one or two characteristic polynomials in the numerator, that is $\mathcal{C}_{1,2}(\mu_{F1};\mu_{B1},\mu_{B2})$ and $\mathcal{C}_{2,2}(\mu_{F1},\mu_{F2};\mu_{B1},\mu_{B2})$, and then treat a special case of the correlation function involving four square roots in the denominator, and two determinants in the numerator, that is  $\mathcal{C}_{2,4}$ in our notation.  As it should be clear from the examples given below the most physically interesting (bulk) scaling regime in  the large-$N$ limit arises when all spectral parameters are close to some value $E\in (-2J,2J)$ by a distance of the order of the mean spacing between neighbouring eigenvalues in the bulk, i.e. $\mathcal{O}(J/N)$. Correspondingly we define the scaled version of the correlation function as
 \begin{equation}\label{1st}
 \mathcal{C}^{(\text{bulk})}_{1,2}(\omega_{F1};\omega_{B1},\omega_{B2})\approx  \left\langle \frac{\det(E+i\omega_{F1}/N-H)}{\det^{1/2}(E+i\omega_{B1}/N-H)\det^{1/2}(E+i\omega_{B2}/N-H)} \right\rangle_{GOE,N \to \infty}
 \end{equation}
 and
\begin{equation}\label{2nd}
  \mathcal{C}^{(\text{bulk})}_{2,2}(\omega_{F1},\omega_{F2};\omega_{B1},\omega_{B2})\approx  \left\langle \frac{\det(E+i\omega_{F1}/N-H)\det(E+i\omega_{F2}/N-H)}{\det^{1/2}(E+i\omega_{B1}/N-H)\det^{1/2}(E+i\omega_{B2}/N-H)} \right\rangle_{GOE,N \to \infty}
 \end{equation}
 where the approximate equality sign above should be understood in the sense of extracting the leading asymptotic dependence on the parameters $\omega_{B}$ and $\omega_F$ when $N\to \infty$. Our results for the above correlation functions are given in Eqs. (\ref{C1same}) and (\ref{C1diff}) for $\mathcal{C}^{(\text{bulk})}_{1,2}$ and in Eqs. (\ref{C2same}) and (\ref{C2diff}) for $\mathcal{C}^{(\text{bulk})}_{2,2}$. In Eq. \eqref{RKab} we provide the result for a special limit (see Eq. \eqref{C4special}) of
 $\mathcal{C}^{\text{(bulk)}}_{2,4}$.  These objects are already rich enough to provide answers for quantities arising in applications of random matrices in the field of Quantum Chaos in closed and open (scattering) systems. We discuss such relations in much detail below.

  Although our methods are specifically tailored for dealing with the GOE we expect our results in the bulk scaling limit to be universal and shared by a broad class of invariant measures on real symmetric matrices $H$ \cite{univinvar} and by so-called Wigner ensembles of random real symmetric matrices with independent, identically distributed entries satisfying relevant moments conditions \cite{univ1,univ2}.

\subsection{Motivations and Background.}

To explain the origin of interest in the correlation functions (\ref{1}) we start with recalling that the phenomenon of Quantum Chaos attracted considerable theoretical and experimental interest for more than three decades and remains one of the areas where applications of Random Matrix Theory are most fruitful and successful \cite{RMTRev}. The applications are based on the famous Bohigas-Giannoni-Schmit (BGS) \cite{BGS} conjecture claiming that in appropriately chosen energy window sequences of highly excited discrete energy levels of generic quantum systems whose classical counterparts are chaotic are statistically indistinguishable from sequences of real eigenvalues of large random matrices of appropriate symmetry. Although not yet fully rigorously proven, this conjecture has an overwhelming support in experimental, numerical and analytical work of the last decades \cite{BGSsemicl}. Inspired by this analogy as well as by the fact of {\it universality} of many random matrix properties (i.e. insensitivity to the particular choice of the probability measure on the matrix space), see \cite{univ1,univ2} and references therein,
one of the common strategies for predicting universal observables of quantum chaotic systems has been expressing them
in terms of resolvents of underlying Hamiltonians, then replacing the actual Hamiltonians by random matrices taken from analytically tractable (usually, Gaussian) ensembles of $N\times N$ random matrices.  The characteristic functions of the  probability densities of the observables under consideration can be frequently computed explicitly by appropriate ensemble averages. Note that the eigenvalues of the standard Gaussian Ensembles, Unitary (GUE, $\beta=2$), Orthogonal (GOE, $\beta=1$) or Symplectic (GSE, $\beta=4$) are independent of the eigenvectors, with the matrix of $N$ orthonormal eigenvectors being  uniformly distributed over the Haar's measure of the Unitary $U(N)$, Orthogonal $O(N)$ or Symplectic $Sp(2N)$ group, correspondingly. To that end it is natural to evaluate the corresponding characteristic functions by performing first the ensemble average over the eigenvectors. For the $\beta=2$ case the average can be frequently done exactly for any $N$ by employing the so-called Itzykson-Zuber-Harish-Chandra \cite{IZ,HC} formula, which is not yet available for $\beta=1,4$ group averages. Nevertheless, one is able to perform the eigenvector averages in the limit $N\gg 1$ by using a heuristic  idea (going back to \cite{Beenakker1994}) that the set of eigenvectors essentially behaves for $N\gg 1$ as if their components were independent, identically distributed Gaussian variables with mean zero and variance $1/N$.
One can rigorously justify this procedure if only a number $n\ll N^{1/2}$ of eigenvectors is involved in the set, see e.g. \cite{eigenvectorstatistics}, but in general a rigorous justification of such a step requires some nontrivial estimates on the resolvents. The heuristic procedure is widely employed in Theoretical Physics for RMT applications to Quantum Chaos using the properties of the standard Gaussian integrals over complex or real variables. In this way the analysis of many distributions of practical interest is reduced to correlation functions of products and ratios involving  {\it  integer} (for $\beta=2,4$) or {\it half-integer} (for $\beta=1$) powers of characteristic polynomials of random matrices. Similar averages arise if one is interested in statistics of the matrix elements of the resolvents computed in the basis of random Gaussian vectors, as it is frequently done in applications to scattering systems with Quantum Chaos, see e.g. the recent paper \cite{FKN13} for an example and further references.
For those and other reasons averages of products and ratios of powers of characteristic polynomials of random matrices attracted much interest over the years. When only {\it integer} powers are involved in the average the corresponding theory was developed for $\beta=2$ in \cite{FS03a,FS03b,BDS} and extended to $\beta=1,4$ in \cite{BS06}. The case of half-integer powers for $\beta=1$ remains however outstanding, despite the fact that it is most relevant for an overwhelming majority of experiments in Quantum Chaos due to the preserved time-reversal invariance of the underlying Hamiltonians. Additional interest to this type of averages gives the fact that they are closely related to the problem of evaluating averages of quantities
involving {\it absolute values} of characteristic polynomials due to the relation  $|\det(E-H)| = \lim_{\epsilon \to 0} \det(E-H+\tfrac{i\epsilon}{N})^{1/2} \det(E-H-\tfrac{i\epsilon}{N})^{1/2}$ valid for matrices $H$ with real eigenvalues. Such averages emerge,  for example, when studying the statistics of the so-called ``level curvatures'' in quantum chaotic systems \cite{mycurv,vOppen1995}, see Eq. (\ref{3}) below, as well as in  the problem of counting the number of stationary points of random Gaussian surfaces, see \cite{Fyo04,my2005}.

To support the above picture we describe below explicitly a few examples of relations between the characteristic functions of the physical observables of interest in quantum chaotic systems which can be related to particular instances of the correlation function (\ref{1}). The list is almost certainly not exhaustive (for example, when writing this article we have learned that the square roots of characteristic polynomials emerged very recently in \cite{Ake14}),  but hopefully representative.
\begin{itemize}
\item  {\bf LDoS distribution.} One of the first examples of that sort which is worth mentioning is related to the statistics of the local density of states (LDoS) $\rho(x;E,\eta)$ at a  point $x$ of a quantum system with  energy levels broadening $\eta$ due to a uniform absorption in the sample. Mathematically the LDoS is defined in terms of the diagonal matrix element of the resolvent as $\rho(x;E,\eta) = \tfrac{1}{\pi}\, \text{Im} \langle x | (E-\tfrac{i\eta}{N}-H)^{-1} | x \rangle$, and one is interested in understanding the statistics of the LDoS assuming a random matrix GOE Hamiltonian $H$ of size $N\times N$,  with the parameter $\eta$ being fixed when $N\to \infty$. The Laplace transform for the probability density ${\cal P}(\rho)$ of the LDoS can be expressed in the large-$N$ limit as \cite{TP1995}
     \begin{equation}\label{2}
        \int_0^{\infty} e^{-s\rho} {\cal P}(\rho)\,d\rho
           =\left\langle \frac{\det^{1/2}\left[(E-H)^2+\frac{\eta^2}{N^2}\right]}{\det^{1/2}\left[(E-H)^2+\frac{\eta^2} {N^2}+\frac{\eta s}{N}\right]} \right\rangle_{GOE,N \to \infty}.
       \end{equation}
 Evaluation of the above random matrix average (which in our notation is a particular case of $\mathcal{C}^{\text{(bulk)}}_{2,4}$ ) attempted in \cite{TP1995} resulted in a quite impractical 5-fold integral, and to this end remains an outstanding RMT problem. Note however that
   the density ${\cal P}(\rho)$ has been found via a different route avoiding (\ref{2}) as a sum of two-fold integrals in \cite{Fyo05,Fyo05rev}.

\item {\bf Probability distribution of ``level curvatures''}. Consider a perturbation $\mathcal{H}=H+\alpha V$ of the Hamiltonian $H$ where $\alpha$ is a control parameter and $V$ is a real symmetric matrix. ``Level curvatures'' are defined as second derivatives of the eigenvalues $\lambda_n(\alpha)$ (interpreted as energy levels of a quantum-chaotic system) with respect to the external parameter $\alpha$: $C_n = \frac{\partial^2 \lambda_n(\alpha)}{\partial\alpha^2} = \sum_{m \neq n} \frac{\langle n|V|m \rangle^2}{\lambda_n-\lambda_m}$.
    Assuming the perturbation $V$ to be taken as well from the GOE one can show that the probability density $P_E(c) =\frac{1}{\bar{\rho}(E)} \left\langle  \sum_{n=1}^N \delta(c-C_n)\delta(E-\lambda_n)\right\rangle$ of the level curvatures for GOE matrices $H$ with eigenvalues $\lambda_n$ and mean density of eigenvalues $\bar{\rho}(E)$ can be represented as\cite{mycurv,vOppen1995}
        \begin{equation}\label{3}
         P_E(c) \propto \int_{-\infty}^{+\infty}\! d\omega\, e^{i\omega c} \left\langle \frac{|\det(E-H)|\det^{1/2}(E-H)}{\det^{1/2}(E+\frac{i\omega}{N}-H)} \right\rangle_{GOE,N \to \infty}
        \end{equation}
        where the required random matrix average in the right-hand side was independently evaluated by several alternative methods in \cite{mycurv,vOppen1995}. Note that heuristic arguments appealing to Gaussianity of GOE eigenvectors in the large-$N$ limit suggest universality of the level curvature distribution for a ``generic'' choice of $V$, and a rigorous proof of this fact is under consideration\cite{AGnotes}.
\item {\bf Statistics of S-matrix poles.} Various questions related to the statistics of quantum chaotic resonances (poles of the scattering matrix in the complex energy plane \cite{FyoSav11}) in the regime of a weakly open scattering system  can be related to evaluation of the averages
 \begin{equation}\label{4}
  \left\langle \frac{\det H^2}{\det^{1/2}(H^2+\frac{\omega^2}{N^2})} \right\rangle_{GOE, N\to \infty}
  \quad \mbox{and} \quad \left\langle \mbox{det}^{1/2}{\left(H^2+\frac{\omega^2}{N^2}\right)} \right\rangle_{GOE,N \to \infty}
 \end{equation}
 where $\omega$ is considered as $N$-independent parameter.
  The first of these averages features in the statistics of resonance widths change
under influence of a small perturbation of the Hamiltonian $H\to H+\alpha V$ akin to that considered above for the level curvature case. Such change reflects the intrinsic non-orthogonality of the associated resonance eigenfunctions \cite{FyoSav12}. Another manifestation of the same non-orthogonality is the statistics of the so-called Petermann factor which again can be related to random matrix averages involving half-integer powers of characteristic polynomials, see \cite{SFPB}. The second average in (\ref{4}) arose in a recent attempt of clarifying the statistics of resonance widths beyond the standard first-order perturbation theory, see  \cite{FyoSav2014}. Evaluating both averages featuring in (\ref{4}) in a uniform way by a systematic procedure was one of our motivations behind writing the present paper.

\item {\bf Statistics of Wigner $K$-matrix.} In the theory of quantum chaotic scattering the Wigner $K$-matrix is essentially defined as a certain projection of the resolvent of $H$. More precisely this is an  $M\times M$ matrix with entries $K_{ab}=W_a^T (E-H)^{-1} W_b$ , with $W_a$ being an $N$-component vector of coupling amplitudes $W_{ia}$ between $N$ energy levels of the  closed system (modelled for a chaotic system by an $N\times N$ random matrix  Hamiltonian $H$) and $M$ scattering channels
   open at a given energy $E$ of incoming waves. Note that the more standard $M\times M$ unitary $S$-matrix is related to $K$ via a simple  Cayley  transform $S=\frac{I-iK}{I+iK}$. In the random matrix approach one usually assumes for the amplitudes $W_{ia}$  either the model of fixed orthogonal channels with $W_a^T W_b=\gamma_a \delta_{ab}$ \cite{VWZ} or independent Gaussian channels where the amplitudes are taken to be i.i.d. Gaussian variables with $\langle W_a^T W_b \rangle = \gamma_a \delta_{ab}$ \cite{Sokolov-Zelevinsky}.

 The quantities $K_{ab}$ are of direct experimental relevance and can  be measured in microwave experiments as they are related to the real part of the electromagnetic impedance \cite{Hem05, Hem06}. For real $E$  in the bulk of the spectrum
the statistics of the diagonal entries $K_{aa}$ is long known to be  given by the same Cauchy distribution for all $\beta=1,2,4$, see e.g. \cite{Fyo97a,FyoWi07}, and very recently was actually shown to be very insensitive to spectral properties of $H$ under rather general conditions \cite{AW13}.
Similarly, one can consider the probability density ${\cal P}(K_{ab})$ of the individual off-diagonal entries $K_{a\ne b}$ for $\beta=1$. For the model of Gaussian channels one arrives to the Fourier transformed ${\cal P}(K_{ab})$ in the form:
\begin{equation}\label{7}
  \int_{-\infty}^\infty e^{ix K_{ab}}{\cal P}(K_{ab})\,dK_{ab}  = \lim_{N\to \infty}\left\langle \frac{|\det(E-H)|}{\det^{1/2}[(E-H)^2+\frac{\gamma_a \gamma_b x^2}{N^2}]} \right\rangle_{GOE}=R_E(x).
\end{equation}
Note that the average featuring in the right-hand side does not follow from either
$\mathcal{C}_{1,2}^{\text{(bulk)}}$ or $\mathcal{C}_{2,2}^{\text{(bulk)}}$ as a special case, but
   is rather a limiting  case of the more general correlation function $\mathcal{C}_{2,4}^{\text{(bulk)}}$ as it can be seen from the following representation:
   \begin{equation}\label{C4special}
R_E(x)=\lim_{\epsilon \to 0}\lim_{N\to \infty}\left\langle \frac{\det^2(E- H)}{\det^{1/2}\left((E-H)^2+\frac{\gamma_a \gamma_b x^2}{N^2}\right)\det^{1/2}\left((E-H)^2+\frac{\epsilon^2}{N^2}\right)} \right\rangle_{GOE}.
\end{equation}

To the best of our knowledge the probability density ${\cal P}(K_{ab})$  for $a\ne b$ (or its Fourier transform) was not yet given explicitly in the literature\footnote{The distribution of the off-diagonal entries $S_{a\ne b}$ of the scattering matrix $S$ is also experimentally relevant \cite{Die10, Soffexp} and has been calculated very recently in \cite{Soff}. However it remains a challenge to extract the statistics of $K_{a\ne b}$ from it in a manageable form.} and we will find it below for the center of the GOE spectrum, see Eq. (\ref{RKab}). Note that it is expected that statistics of the $K$-matrix entries for a GOE Hamiltonian $H$ is the same for the two choices of the coupling $W$ as long as $M$ stays finite for $N\to \infty$.

As to the $M\times M$  matrix $K$ as a whole,  the probability density $\mathcal{P}(K)$  for $\beta=1$ and $E$ in the bulk of the spectrum is expected to be given by a Cauchy-like expression:
        \begin{equation}\label{5}
         \mathcal{P}(K) \propto \det[\lambda^2+(K-\langle K \rangle)^2]^{-\frac{M+1}{2}}
\end{equation}
with $E$-dependent mean $\langle K \rangle$ and the width parameter $\lambda$. This distribution was conjectured in 1995 by P. Brouwer  on the experience of working with $H$ from the so-called Lorentzian ensemble, see \cite{Bro95}.
 A similar formula for invariant ensembles of complex Hermitian random matrices $H$ ( i.e. $\beta=2$) was proved rigorously very recently in \cite{FKN13}, and in the same paper   it was mentioned that for $\beta=1$ and the case of random Gaussian coupling the following relation holds\footnote{The corresponding formula in \cite{FKN13} was written not accurately enough and did not show the dependence on $\sgn\det$ factors.}:
        \begin{equation}\label{6}
     \int\! e^{i \text{Tr}(KX)}   \mathcal{P}(K) dK = \lim_{N \to \infty}\left \langle \prod_{c=1}^M \frac{\det^{1/2}(E-H)\left[ \sgn \det(E-H)\right]^{\Theta(-x_c)}}{\det^{1/2}(E+\frac{i\gamma_c x_c}{N}-H)} \right \rangle_{GOE}
        \end{equation}
        where $\Theta(-x_c)=1$ for negative $x_c$ and is zero otherwise. Although our attempts to verify Brouwer's
        conjecture for $\beta=1, M=2$ along these lines were not fully successful yet, we discuss
        partial results, see (\ref{M=2a})-(\ref{M=2b}) below.

 \item A particular type of the correlation functions (\ref{1}) was investigated in \cite{FK2003} where it has been shown that for any integer $k>0$ and fixed real $\delta$ holds \footnote{Note also that an ensemble average closely related to the left-hand side of (\ref{fyokeat}) was evaluated explicitly in \cite{ForrKeat}, with the general circular $\beta-$ensemble  replacing the GOE. The result was expressed for all $\beta>0$ and all integer $N\ge 1$ in terms of a certain generalised hypergeometric function. The $\delta\to 0$ asymptotics for large $N\gg 1$ of the latter function does agree with the one following from the right-hand side of (\ref{fyokeat}).}
 \begin{equation}\label{fyokeat}
  \begin{split}
 &\left\langle \frac{1}{\det^{k/2}(i\delta/N-H)\det^{k/2}(-i\delta/N-H)} \right\rangle_{GOE,N \to \infty} \\
 &\propto e^{k\delta}\int_1^{\infty}\frac{d\lambda_1e^{-\delta\lambda_1}}{\sqrt{\lambda_1^2-1}}\ldots \int_1^{\infty}\frac{d\lambda_ke^{-\delta\lambda_k}}{\sqrt{\lambda_k^2-1}}\,\prod_{i<j}^k|\lambda_i-\lambda_j|.
 \end{split}
 \end{equation}

\end{itemize}

\subsection{The results.}
\begin{itemize}
\item As it has been mentioned above, we were not yet able to reveal nice mathematical structures for (\ref{1}) at finite values of the matrix size $N$ beyond the simplest case $K=1, L=1$, where the methods outlined below yielded a determinantal structure which we give here for completeness:
 \begin{equation}\label{8}
\begin{split}
  \mathcal{C}_{1,1}(\mu_F;\mu_B)=&
\left(\frac{J^2}{2N}\right)^{N/4} \frac{[-i\sgn(\im(\mu_B))]^{N+1}}{ \Gamma(N/2)} \\
&\times \det{\left(\begin{array}{cc} H_{N-1}\left(\frac{\sqrt{N}}{J}\mu_F\right) &  F_{N/2-1}\left(\frac{\sqrt{N}}{\sqrt{2}J}\mu_B\right)\\ H_{N}\left(\frac{\sqrt{N}}{J}\mu_F\right) &  F_{N/2}\left(\frac{\sqrt{N}}{\sqrt{2}J}\mu_B\right)\end{array}\right)}
\end{split}
 \end{equation}
where $\Gamma(x)$ is the Euler Gamma-function, $H_{N}(z)=\frac{i^N}{\sqrt{2\pi}}\int_{-\infty}^{\infty}dt\, t^N \exp[-\tfrac{1}{2}(t+iz)^2]$ is a Hermite polynomial and the function
\[
F_{N}(z)=[i \sgn(\im(z))]^N\int_0^{\infty}dt\, t^N\exp[-\tfrac{1}{2}(t^2+2i \sgn(\im(z))z t) ]
\]
may be associated with the Cauchy transforms of Hermite polynomials \cite{FS03a}.

\item The explicit forms for the ``bulk'' correlation functions $\mathcal{C}^{(\text{bulk})}_{1,2}(\omega_{F1};\omega_{B1},\omega_{B2})$ (see Eq. \eqref{1st}) and $\mathcal{C}^{(\text{bulk})}_{2,2}(\omega_{F1},\omega_{F2};\omega_{B1},\omega_{B2})$ (see Eq. \eqref{2nd}) depend very essentially on the signs of $\omega_{B1}$ and $\omega_{B2}$. In particular, if $\sgn \omega_{B1}=\sgn \omega_{B2}$ the first correlation function is given by
\begin{equation} \label{C1same}
 \mathcal{C}^{(\text{bulk, } \sgn \omega_{B1}=\sgn \omega_{B2})}_{1,2}(\omega_{F1};\omega_{B1},\omega_{B2})\approx e^{\frac{2\omega_{F1}-\omega_{B1}-\omega_{B2}}{4J^2}(iE+\sgn{\omega_B} \sqrt{4J^2-E^2})},
\end{equation}
whereas for $\sgn \omega_{B1}=-\sgn \omega_{B2}$ the same object takes instead the form
\begin{equation}
\begin{split} \label{C1diff}
  &\mathcal{C}^{(\text{bulk, }\sgn \omega_{B1}=-\sgn \omega_{B2})}_{1,2}(\omega_{F1};\omega_{B1},\omega_{B2}) \approx
\frac{(-i)^N}{\pi\sqrt{2 N \rho}(2J)^{N+1}}
\,e^{-\frac{iE}{4J^2}(\omega_{B1}+\omega_{B2}-2\omega_{F1})}  \\
&\times\bigg\{[Ae^{-\pi \rho\omega_{F1}} -(-1)^N A^*e^{+\pi  \rho\omega_{F1}} ]
(\omega_{B1}+\omega_{B2}-2\omega_{F1})K_0\left(\tfrac{\pi \rho}{2}|\omega_{B1}-\omega_{B2}|\right) \\
& \phantom{\times \bigg\{}+[Ae^{-\pi \rho\omega_{F1}}+(-1)^N A^*e^{+\pi \rho\omega_{F1}}]
|\omega_{B1}-\omega_{B2}|K_1\left(\tfrac{\pi \rho}{2}|\omega_{B1}-\omega_{B2}|\right) \bigg\}
\end{split}
\end{equation}
with
\begin{equation} \label{A}
 A(E,N)=(2\pi J^2 \rho+iE)^{N-1/2}\ e^{\frac{i\pi N}{2}\rho E},
\end{equation}
where we introduced $\rho=\frac{1}{2\pi J^2}\sqrt{4J^2-E^2}$ for the mean eigenvalue density of large GOE matrices in the bulk of the spectrum and used the standard notation $K_m(z)$ for the modified Bessel (Macdonald) functions of second kind and index $m$. Note that the asymptotic expression (\ref{C1diff}) shows an interesting ``parity effect'': it behaves differently depending on whether $N$ is even or odd for arbitrary large values of $N$.

Similarly the second correlation function for $\sgn \omega_{B1}=\sgn \omega_{B2}$ is given by
\begin{equation}
 \begin{split} \label{C2same}
   &\mathcal{C}^{(\text{bulk, }\sgn \omega_{B1}=\sgn \omega_{B2})}_{2,2}(\omega_{F1},\omega_{F2};\omega_{B1},\omega_{B2})\approx \\
 &\left(\frac{J}{\sqrt{N}}\right)^N \frac{3\tilde{H}_N\left(\frac{\sqrt{N}E}{J}\right)}{[\pi\rho(\omega_{F1}-\omega_{F2})]^3} e^{\frac{iE(\omega_{F1}+\omega_{F2})}{2J^2}}\,
e^{-\frac{iE(\omega_{B1}+\omega_{B2})}{4J^2}} e^{-\frac{\pi \rho(|\omega_{B1}|+|\omega_{B2}|)}{2}} \\
&\times \left[\pi  \rho(\omega_{F1}-\omega_{F2})\cosh\left(\pi \rho(\omega_{F1}-\omega_{F2})\right)
-\sinh\left(\pi\rho(\omega_{F1}-\omega_{F2})\right) \right],
 \end{split}
\end{equation}
where $\tilde{H}_N\left(\frac{\sqrt{N}E}{J}\right)=\sqrt{2}\left(\frac{iN}{2J}\right)^N e^{-N/2}\, e^{\frac{N}{4J^2}E^2} [(-1)^N A(E,N)+ A^*(E,N)]$ is the appropriate large-$N$ asymptotic of the $N$-th Hermite polynomial, with $A(E,N)$ defined in Eq. \eqref{A}.  In the case $\sgn \omega_{B1}=-\sgn \omega_{B2}$ we get instead
\begin{equation}
\begin{split} \label{C2diff}
 &\mathcal{C}^{(\text{bulk, }\sgn \omega_{B1}=-\sgn \omega_{B2} )}_{2,2}(\omega_{F1},\omega_{F2};\omega_{B1},\omega_{B2}) \approx \\
&\sqrt{\frac{2N}{\pi}} \frac{J^{N+1} e^{-N/2}}{(\omega_{F1}-\omega_{F2})^3} e^{\frac{N}{4J^2}E^2} e^{\frac{iE(\omega_{F1}+\omega_{F2})}{2J^2}}\,
e^{-\frac{iE(\omega_{B1}+\omega_{B2})}{4J^2}}  \\
& \bigg\{ [(\omega_{F1}+\omega_{F2})(\omega_{B1}+\omega_{B2})-2\omega_{F1}\omega_{F2}-2\omega_{B1}\omega_{B2}] K_0 \left(\tfrac{\pi \rho}{2}|\omega_{B1}-\omega_{B2}| \right)  \\
&\hskip1em\times \left[\pi \rho(\omega_{F1}-\omega_{F2})\cosh\left(\pi \rho(\omega_{F1}-\omega_{F2})\right)
-\sinh\left(\pi \rho(\omega_{F1}-\omega_{F2})\right) \right]   \\
& \hskip0.5em + \pi \rho(\omega_{F1}-\omega_{F2})^2 |\omega_{B1}-\omega_{B2}| \sinh\left(\pi \rho(\omega_{F1}-\omega_{F2})\right)
K_1 \left(\tfrac{\pi \rho}{2}|\omega_{B1}-\omega_{B2}| \right) \bigg\}.
\end{split}
\end{equation}
Note that the parity of $N$ plays no role for the large-$N$ behaviour of this correlation function.
\end{itemize}
Let us now discuss a few special cases motivated by applications mentioned above.
\begin{itemize}
\item The characteristic function of the ``level curvatures'', Eq. \eqref{3} can be represented as a special limit of $\mathcal{C}^{(\text{bulk})}_{2,2}$,
\begin{equation}
\begin{split}
 \left\langle\frac{|\det(E-H)|\det(E-H)^{1/2}}{\det(E+i\omega/N-H)^{1/2}} \right\rangle_{GOE,N\to \infty}
&=\lim_{\epsilon \to 0}\mathcal{C}^{(\text{bulk})}_{2,2}(\epsilon,-\epsilon;-\epsilon,\omega) \\
&\propto e^{-\frac{i E}{4J^2}\omega}|\omega| K_1\left(\tfrac{\sqrt{4J^2-E^2}}{4J^2}|\omega| \right).
\end{split}
\end{equation}
The Fourier transform of this result (for brevity we choose $E=0$, $J=1$) yields the curvature distribution,
\begin{equation}
 P(c) = \frac{1}{4\pi} \int_{-\infty}^\infty d\omega |\omega| K_1\left(\tfrac{1}{2}|\omega|\right) \exp(i\omega c) = (1+4c^2)^{-3/2},
\end{equation}
which coincides with the expression found in earlier works by alternative methods \cite{mycurv,vOppen1995}.

\item The two averages featuring in Eq. \eqref{4} can be recovered as special cases from $\mathcal{C}^{(\text{bulk})}_{2,2}$ and are for the choice $J=1$ given by
\begin{equation}
\begin{split}
  &\left\langle \frac{\det^2 H}{\det^{1/2}(H^2+\tfrac{\omega^2}{N^2})} \right\rangle_{GOE, N \to \infty}= \mathcal{C}^{(\text{bulk})}_{2,2}(0,0;\omega,-\omega) \\
  &\approx  2\sqrt{\frac{2N}{\pi}}e^{-N/2} \left[ \frac{\omega^2}{3} K_0 \left(|\omega| \right) +  |\omega| K_1 \left(|\omega| \right) \right],
\end{split}
 \end{equation}
\begin{equation} \label{eqscatt2}
\begin{split}
  & \left\langle \det(H^2+\tfrac{\omega^2}{N^2})^{1/2} \right\rangle_{GOE,N\to \infty}=\mathcal{C}^{(\text{bulk})}_{2,2}(\omega,-\omega;\omega,-\omega) \\
  &\hskip1em\approx   \sqrt{\frac{2N}{\pi}}e^{-N/2} \bigg[ \left(\cosh(2\omega) -\frac{\sinh(2\omega)}{2\omega} \right)  K_0(|\omega|)+  \sinh(2|\omega|) K_1 (|\omega|) \bigg].
\end{split}
 \end{equation}
The above formulas have been already presented in \cite{FyoSav12,FyoSav2014}, with derivation relegated to the present paper. We tested the validity of (\ref{eqscatt2}) by direct numerical simulations of GOE matrices of a moderate size, see figure \ref{scatt2}.
\begin{figure}[!h]
\centering
\includegraphics[width=0.5\textwidth]{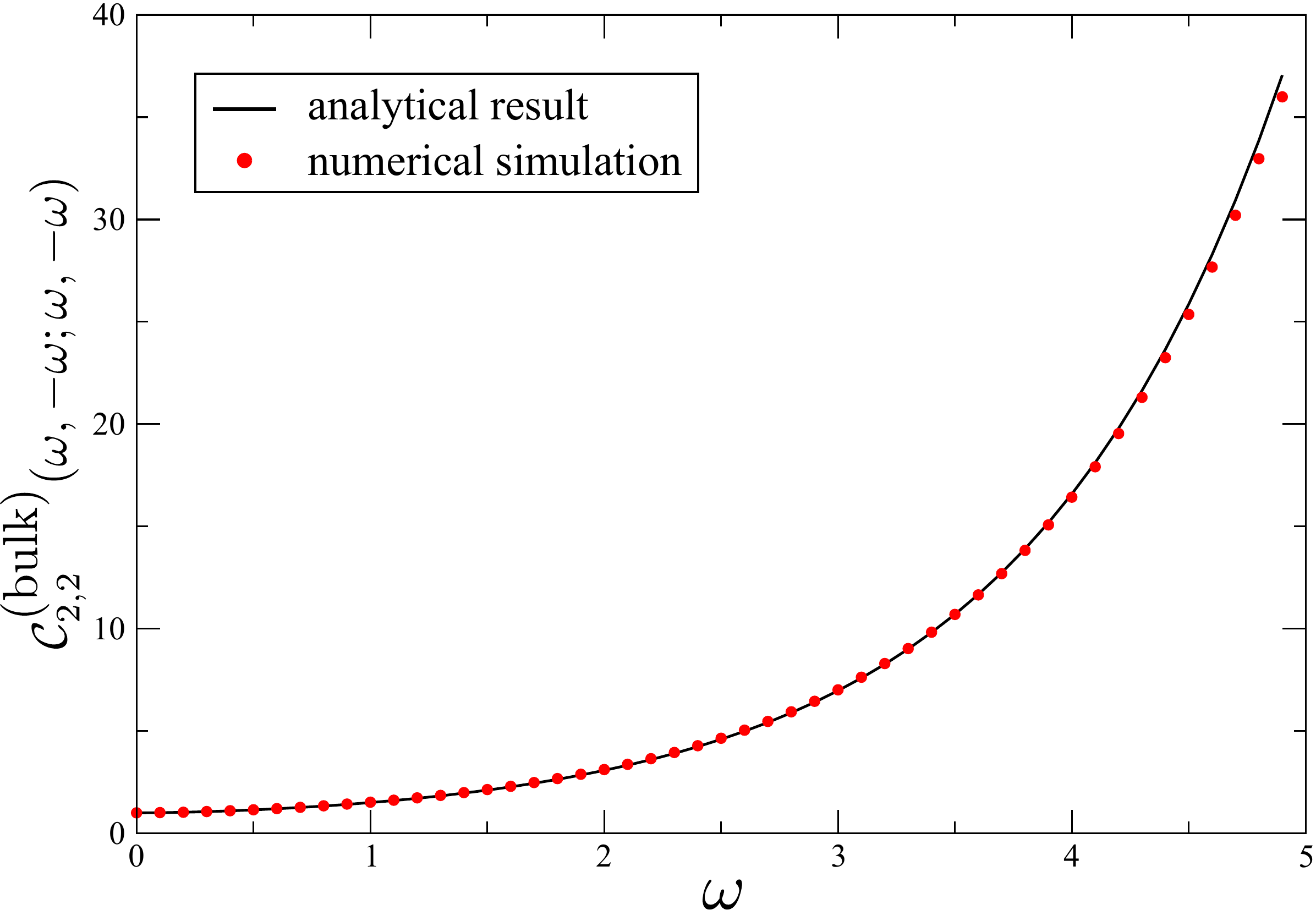}
\caption{The correlation function $\mathcal{C}^{(\text{bulk})}_{2,2}(\omega,-\omega;\omega,-\omega)$ from Eq. \eqref{eqscatt2} against numerical results obtained from a sample of 40000 GOE-matrices of size $80 \times 80$. }
\label{scatt2}
\end{figure}

\item For the characteristic function of an off-diagonal element $K_{ab}$ of the $K$-matrix, see Eq. \eqref{7}, we choose to present the corresponding result only for the so-called ``perfect coupling'' case, i.e. $E=0$ and $\gamma_a = \gamma_b =1$, the case of general $\gamma_a\ne \gamma_b$ following by a trivial rescaling. It is given by
   \begin{equation} \label{RKab}
    \lim_{N \to \infty}\left\langle \frac{|\det H|}{\det(H^2+\frac{x^2}{N^2})^{1/2}} \right\rangle_{GOE}
     =\frac{2}{\pi}\left(\frac{|x|}{J} K_0(|x|/J)+\int_{|x|/J}^\infty\!dy\, K_0(y) \right).
   \end{equation}
The ensuing distribution $\mathcal{P}(K_{ab})$ is then consequently given by its Fourier transform,
\begin{equation} \label{PKab}
 \mathcal{P}(K_{ab}) =  \frac{2}{\pi^2(1+K_{ab}^2)} \left(1+\frac{\text{arsinh}(K_{ab})}{K_{ab}\sqrt{1+K_{ab}^2}} \right).
\end{equation}
 In the  \ref{app1} we verify that this result is in complete agreement with Brouwer's conjecture
 claiming that $K$ for the ``perfect coupling'' case is distributed as $\mathcal{P}(K) \propto \det[1+K^2]^{-(M+1)/2}$.  We also check these expressions against direct numerical simulations, see figure
 \ref{PandRKab}.

\begin{figure}[!h]
\centering
\includegraphics[width=0.9\textwidth]{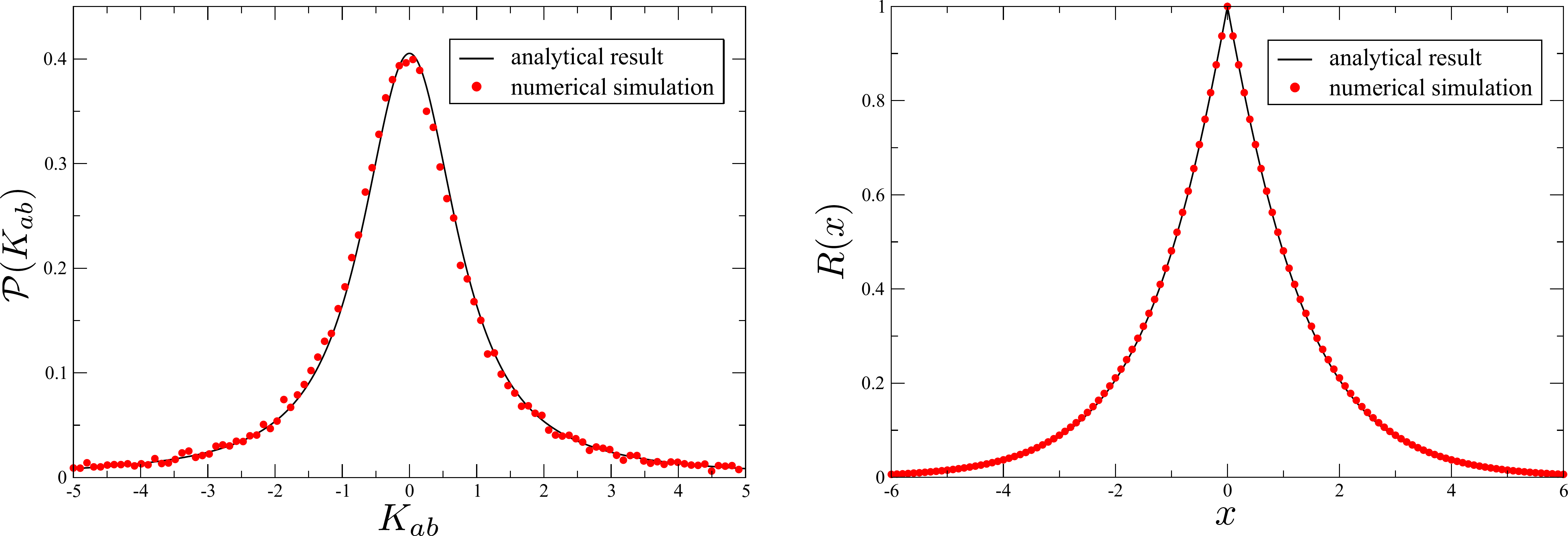}
\caption{Distribution of an off-diagonal $K$-matrix element $K_{ab}$ (left) and its characteristic function (right). The numerical results were obtained from samples of 40000 GOE-matrices of size $80 \times 80$. }
\label{PandRKab}
\end{figure}

\item The $M=2$ case of Eq. \eqref{6} features the correlation function
        \begin{equation}\label{M=2a}
        \left \langle \frac{\det(E-H) \sgn \det(E-H)^{\Theta(-x_1 x_2)}}{\det^{1/2}(E+\frac{i\gamma_1 x_1}{N}-H) \det^{1/2}(E+\frac{i\gamma_2 x_2}{N}-H)} \right \rangle_{GOE}.
        \end{equation}
Assume that $x_1 x_2>0$ so that $\Theta(-x_1 x_2)=0$ and the sign-factor is immaterial. The correlation function then takes the form of
\begin{equation}
\mathcal{C}^{(\text{bulk})}_{1,2}(0;\gamma_1 x_1,\gamma_2 x_2)\approx e^{\frac{-\gamma_1 x_1-\gamma_2 x_2}{4J^2}(iE+\sgn x_1 \sqrt{4J^2-E^2})},
\end{equation}
which simplifies even further to $e^{\frac{- |x_1|- |x_2|}{2J}}$ for the ``perfect coupling'' case $E=0$, $\gamma_1=\gamma_2=1$. In the opposite case $x_1 x_2 <0$ on the other hand the correlation function takes the form
        \begin{equation}\label{M=2b}
        \left \langle \frac{|\det(E-H)|}{\det^{1/2}(E+\frac{i\gamma_1 x_1}{N}-H) \det^{1/2}(E+\frac{i\gamma_2 x_2}{N}-H)} \right \rangle_{GOE},
        \end{equation}
which is again a special case of $\mathcal{C}^{(\text{bulk})}_{2,4}$. In the particular case $\gamma_1 x_1=-\gamma_2 x_2 \equiv \gamma x$, the above expression assumes the same form as one needed for extracting the distribution of a single off-diagonal element $K_{ab}$, see Eq. \eqref{7} and \eqref{RKab}. While a full proof that $K$ is distributed according to the Cauchy distribution, Eq. \eqref{5}, requires the knowledge of the above expression for arbitrary values of $x_1$ and $x_2$,  one can show  that our partial results for $\gamma_1 x_1=-\gamma_2 x_2 \equiv \gamma x$ are indeed consistent with  Eq. \eqref{5}, see \ref{app2}.

\item Finally we notice that an interesting special case of $\mathcal{C}^{(\text{bulk})}_{1,2}$ is the average of the sign of the GOE characteristic polynomial given asymptotically by
\begin{equation}
\begin{split}
 \langle \sgn \det(E-H) \rangle_{GOE, N\to \infty} &=\lim_{\epsilon \to 0}\mathcal{C}^{(\text{bulk})}_{1,2}(0;\epsilon,-\epsilon) \\
&\approx  \frac{2J^2(-i/(2J))^N}{\sqrt{\pi N}(4J^2-E^2)^{3/4}} [A(E,N)+(-1)^N A^*(E,N)],
\end{split}
\end{equation}
where $A(E,N)$ is defined in Eq. \eqref{A}.
\end{itemize}

\section{Derivation of the main results.}
\subsection{Evaluation of the correlation functions Eq.(\ref{1st}) and Eq.(\ref{2nd}).}
  At present the only systematic method for evaluating the ensemble averages $ \mathcal{C}^{(\text{bulk})}_{1,2}$ and  $\mathcal{C}^{(\text{bulk})}_{2,2}$ seems to be the so-called supersymmetric formalism, see \cite{Guhroxbook} and references therein. Within RMT framework several variants of that method  are by now well-developed and we will follow one of them proposed in \cite{Fyo2002}.  We only outline the major steps of the procedure below referring the interested reader to the cited literature and leaving technical detail
   for  \cite{thesis}. To that
end we start with replacing the square roots of determinants in the denominator by Gaussian integrals over $N$-component real vectors ${\bf x}_i$, and the determinants in the numerator by integrals over vectors $\zeta_i$ whose $N$ components are complex anticommuting (Grassmann) variables. In that way the correlation function $\mathcal{C}^{(\text{bulk})}_{1,2}$ can be represented by
\begin{equation}
\begin{split}
 \mathcal{C}^{(\text{bulk})}_{1,2} \propto \Bigg\langle &\int d{\bf x}_1\, e^{\frac{i}{2} s_1 {\bf x}_1^T(E+i\omega_{B1}/N-H){\bf x}_1}
     \int d{\bf x}_2\, e^{\frac{i}{2} s_2 {\bf x}_2^T(E+i\omega_{B2}/N-H){\bf x}_2}  \\
& \times \int d^2\zeta\, e^{-\frac{i}{2} \zeta^{\dag}(E+i\omega_{F1}/N-H)\zeta} \Bigg\rangle,
\end{split}
\end{equation}
and similarly for $\mathcal{C}^{(\text{bulk})}_{2,2}$ where we have to introduce one more integration over a vector of $N$ anticommuting components. Note that we have to introduce $s_i \equiv \sgn \omega_{Bi}$ in order to render the integrals over the commuting variables convergent.

The ensemble average can now easily be performed and yields for $\mathcal{C}^{(\text{bulk})}_{2,2}$ the result,
\begin{equation}
 \langle e^{-\frac{i}{2} [s_1 {\bf x}_1^T H {\bf x}_1 + s_2 {\bf x}_2^T H {\bf x}_2 - \zeta_1^\dag H \zeta_1- \zeta_2^\dag H \zeta_2]} \rangle
  =e^{-\frac{J^2}{4N}[\tr (Q_B L)^2 -\tfrac{1}{2}\tr Q_F^2 +\zeta_1^T \zeta_2 \zeta_2^\dag \zeta_1^*-2 \zeta_1^\dag B \zeta_1 -2 \zeta_2^\dag B \zeta_2]},
\end{equation}
where we introduced the $N\times N$ matrix $B=s_1 {\bf x}_1 \otimes {\bf x}_1^T + s_2 {\bf x}_2 \otimes {\bf x}_2^T$ as well as the $2\times 2$  matrices
\begin{equation}
 Q_F=\begin{bmatrix}
      \zeta_1^\dag \zeta_1 & \zeta_1^\dag \zeta_2 \\
      \zeta_2^\dag \zeta_1 & \zeta_2^\dag \zeta_2
     \end{bmatrix},\
 Q_B=\begin{bmatrix}
      {\bf x}_1^T {\bf x}_1 & {\bf x}_1^T {\bf x}_2 \\
      {\bf x}_2^T {\bf x}_1 & {\bf x}_2^T {\bf x}_2
     \end{bmatrix},\
 L=\begin{bmatrix}
      s_1 & 0 \\
      0 & s_2
     \end{bmatrix}.
\end{equation}
 A similar expression for $\mathcal{C}^{(\text{bulk})}_{1,2}$ can be obtained from the above by replacing all terms containing $\zeta_2$ with 0 so that $Q_F$ becomes a scalar in this case. At the next step we employ a Hubbard-Stratonovich transformation for the anticommuting variables only by exploiting the identity
\begin{equation}
 \exp \left(\frac{J^2}{8N} \tr Q_F^2 \right) \propto \int d\widehat{Q}_F\, \exp \left(-\tr \widehat{Q}_F^2+\frac{J}{\sqrt{2N}} \tr \widehat{Q}_F Q_F \right),
\end{equation}
where $\widehat{Q}_F=\begin{bmatrix} q_{11} & q_{12} \\ q_{12}^* & q_{22} \end{bmatrix}$ is a Hermitian $2\times 2$ matrix of commuting variables for $\mathcal{C}^{(\text{bulk})}_{2,2}$ and a single scalar variable $\widehat{Q}_F \equiv q$  for $\mathcal{C}^{(\text{bulk})}_{1,2}$. For $\mathcal{C}^{(\text{bulk})}_{2,2}$ we also need to bilinearise the term $\zeta_1^T\zeta_2\zeta_2^\dag\zeta_1^*$ which can be achieved by introducing an auxiliary Gaussian integral over a complex variable $u$, with $u^*$ standing for its conjugate:
\begin{equation}
 \exp \left(-\frac{J^2}{4N}\zeta_1^T\zeta_2\zeta_2^\dag \zeta_1^* \right) =\int d^2 u\, \exp\left(-u^* u - \frac{iJ}{2\sqrt{N}}(u \zeta_1^\dag \zeta_2^*+u^* \zeta_2^T \zeta_1)\right).
\end{equation}
With the integrand being bilinear in the Grassmann vectors it is easy to perform the integration over the anticommuting variables explicitly. The resulting expression in both cases depends on the ${\bf x}$-vectors only via the eigenvalues of the matrix $Q_BL$. This allows us to follow  the route explained in detail in \cite{FK2003,Fyo2002}  and to employ the identity from Appendix D of \cite{FS02}:
\begin{equation} \label{IT}
 \int d{\bf x}_1 \dots \int d{\bf x}_n\, F(Q_B) \propto \int_{\widehat{Q}_B>0} d\widehat{Q}_B (\det \widehat{Q}_B)^{\frac{N-n-1}{2}} F(\widehat{Q}_B),
\end{equation}
 which helps one to  replace the integration over $n$ real vectors of dimension $N$ by an integral over a positive definite real symmetric matrix $\widehat{Q}_B$ of dimension $n \times n$, where in both our cases actually $n=2$. In the first case this procedure leads us after a trivial rescaling of the integration variables to
 \begin{equation}
\begin{split} \label{c1withQ}
 \mathcal{C}^{(\text{bulk})}_{1,2} \propto &\int_{-\infty}^\infty dq\,q^{N-2}\, e^{-\frac{N}{2J^2}(q-iE+\frac{\omega_F}{N})^2} \int_{Q_B>0} dQ_B (\det Q_BL)^{\frac{N-3}{2}} \det\left(q-Q_BL \right) \\
  &\times e^{-\frac{N}{4J^2}\tr(Q_BL)^2} e^{\frac{iN}{2J^2} \tr Q_BL M_B},
\end{split}
\end{equation}
where for notational convenience we omitted the hats here and henceforth. Similarly, in the second case we arrive at
 \begin{equation}
\begin{split} \label{c2withQ}
 \mathcal{C}^{(\text{bulk})}_{2,2} \propto  &\int_{Q_B>0} dQ_B (\det Q_B L)^{\frac{N-3}{2}}\, e^{-\frac{N}{4J^2}\tr (Q_B L)^2+\frac{iN}{2J^2}\tr Q_B L M_B} \int d^2 u\, e^{-\frac{N}{J^2}u^* u} \\
&\times\int dQ_F\, e^{-\frac{N}{2J^2}\tr Q_F^2-\frac{iN}{J^2}\tr Q_F M_F +\frac{N}{2J^2}\tr M_F^2}  \left[\det Q_F -u^* u\right]^{N-2} \\
& \times \prod_{j=1}^2 \left[\det Q_F-u^* u +\lambda_B^{(j)} \tr Q_F + (\lambda_B^{(j)})^2 \right].
\end{split}
\end{equation}
Here we introduced the $2\times 2$ matrices $M_{B(F)}=E 1_2+\frac{i}{N} \diag(\omega_{B1(F1)},\omega_{B2(F2)})$ and used $\lambda_{B1}$ and $\lambda_{B2}$ for the real eigenvalues of the $2\times 2$ non-selfadjoint matrix $Q_B L$, see
\cite{FK2003,Fyo2002} for technical details.

Setting aside the issue of performing the integration over the matrix $Q_B$ for the time being, in the first case the procedure leaves us with a single $q$-integration whereas in the second case we have to deal with an integral over the $2\times 2$ Hermitian matrix $Q_F$ which contains four independent variables,  and in addition with  integrals over the complex variable $u$. To simplify the integrand we then use that $Q_F$ can be diagonalised by a unitary transformation $Q_F = U \diag(q_{F1},q_{F2}) U^\dag$. The integration over the unitary group can then be performed using the Itzykson-Zuber-Harish-Chandra (IZHC) formula\cite{IZ,HC} which reduces the integration variables to the set  $q_{F1}$, $q_{F2}$, $u$ and $u^*$. Next we note that by introducing a matrix $R=\begin{bmatrix} q_{F1} & u \\ u^* & q_{F2} \end{bmatrix}$ one can express the integrand in terms of $R$ (note e.g. that $\det Q_F -u^* u = \det R$, $\tr Q_F^2+2u^*u = \tr R^2$ etc.). This latter matrix is Hermitian as well, so can also be diagonalized by a unitary transformation $R=U_2 \diag(r_1,r_2)U_2^\dag$.  Although the group integral is not of the IZHC type in this case, it still can be performed explicitly. Following this procedure the correlation function simplifies to
\begin{equation}
\begin{split} \label{c2withQB}
 \mathcal{C}^{(\text{bulk})}_{2,2} \propto  &\frac{e^{\frac{N}{2J^2}\tr M_F^2}}{(\omega_{F1}-\omega_{F2})^3}
\int_{Q_B >0} dQ_B (\det Q_B L)^{\frac{N-3}{2}}\, e^{-\frac{N}{4J^2}\tr (Q_B L)^2+\frac{iN}{2J^2}\tr Q_B L M_B} \\
&\times\int_{-\infty}^\infty dr_1 \int_{-\infty}^\infty dr_2\, e^{-\frac{N}{2J^2}(r_1^2+r_2^2)-\frac{iNE}{J^2}(r_1+r_2)+\frac{1}{2J^2}(r_1+r_2)(\omega_{F1}+\omega_{F2})}  \\
& \times (r_1-r_2)(r_1 r_2)^{N-2} (r_1+\lambda_B^{(1)})(r_2+\lambda_B^{(1)})(r_1+\lambda_B^{(2)})(r_2+\lambda_B^{(2)}) \\
&\times \left[\frac{(r_1-r_2)(\omega_{F1}-\omega_{F2})}{2J^2}\cosh\left(\frac{(r_1-r_2)(\omega_{F1}-\omega_{F2})}{2J^2}\right) \right.\\
&\phantom{\times\bigg[}\left. -\sinh\left(\frac{(r_1-r_2)(\omega_{F1}-\omega_{F2})}{2J^2}\right) \right].
\end{split}
\end{equation}
 At the final step we aim at simplifying the integral over $Q_B$, which in both cases is a $2 \times 2$ real symmetric positive definite  matrix. As the integrands in \eqref{c1withQ} and \eqref{c2withQB} actually depend on the combination $Q_B L$ we change the integration from $Q_B$ to $Q_B L$. Recall that the matrix $L=\diag(\sgn \omega_{B1},\sgn \omega_{B2})$ reflects the signs of $\omega_{B1}$ and $\omega_{B2}$ and this fact will play now a crucial role. If $\omega_{B1}$ and $\omega_{B2}$ are of the same sign, $L$ is proportional to the identity  and hence $Q_B L$ is still positive definite real symmetric and can be diagonalized by an orthogonal transformation $Q_B L = \pm O \diag(p_1,p_2) O^T$. If, however, the signs are different (we may assume for definiteness $\omega_{B1}>0$ and $\omega_{B2}<0$), then the matrix $Q_B L$ will have an underlying hyperbolic symmetry and can be parametrised as \cite{FK2003,Fyo2002}
\begin{equation}
 Q_B L = \begin{bmatrix}
       \frac{p_1-p_2}{2}+\frac{p_1+p_2}{2}\cosh\theta & \frac{p_1+p_2}{2}\sinh\theta \\
       -\frac{p_1+p_2}{2}\sinh\theta & \frac{p_1-p_2}{2}-\frac{p_1+p_2}{2}\cosh\theta
      \end{bmatrix},
\end{equation}
where $p_1,p_2>0$ and $\theta \in (-\infty,\infty)$.
The only term in the integrands \eqref{c1withQ} and \eqref{c2withQB} which actually depends on $\theta$ is $\tr Q_B L M_B = E(p_1-p_2)+\frac{i}{2N}[(p_1-p_2)(\omega_{B1}+\omega_{B2})+(p_1+p_2)(\omega_{B1}-\omega_{B2})\cosh(2\theta)]$. For the $\sgn \omega_{B1}=\sgn \omega_{B2}$ case one obtains the same type of expression with $p_2 \to -p_2$ and $\cosh(2\theta) \to \cos(2\theta)$. The $\theta$-integration can be performed explicitly using
\begin{align}
  &\int_{0}^{2\pi} d\theta\, e^{-\frac{\omega_{B1}-\omega_{B2}}{4J^2}(p_1-p_2)\cos(2\theta)} = 2\pi I_0\left(\frac{\omega_{B1}-\omega_{B2}}{4J^2}(p_1-p_2) \right), \label{BesselI} \\
  &\int_{-\infty}^{\infty} d\theta\, e^{-\frac{\omega_{B1}-\omega_{B2}}{4J^2}(p_1+p_2)\cosh(2\theta)} = 2 K_0\left(\frac{\omega_{B1}-\omega_{B2}}{4J^2}(p_1+p_2) \right), \label{BesselK}
\end{align}
where $I_0(x)$ and $K_0(x)$ stand for the modified Bessel function of the first and second kind, respectively. in this way we arrive at the final expression which is exact for arbitrary value of $N$,
\begin{equation}
\begin{split} \label{C12exact}
 \mathcal{C}^{(\text{bulk},+-)}_{1,2} \propto & \int_{-\infty}^{\infty} dq\, q^{N-2}\,e^{-\frac{N}{2J^2}(q-iE)^2-\frac{\omega_F}{J^2}(q-iE)-\frac{\omega_F^2}{2NJ^2}} \\
&\times \int_0^\infty dp_1  \int_0^\infty dp_2\, (p_1 p_2)^{\frac{N-3}{2}}\, e^{-\frac{N}{4J^2}(p_1^2+p_2^2)+\frac{iNE}{2J^2} (p_1-p_2)-\frac{\omega_{B1}+\omega_{B2}}{4J^2}(p_1-p_2)} \\
  &\times K_0 \left( \frac{(\omega_{B1}-\omega_{B2})(p_1+p_2)}{4J^2} \right) (q-p_1)(q+p_2)(p_1+p_2).
\end{split}
\end{equation}
 and
\begin{equation}
\begin{split} \label{C22exact}
\mathcal{C}^{(\text{bulk},+-)}_{2,2} \propto &\frac{e^{\frac{N}{2J^2}\tr M_F^2}}{(\omega_{F1}-\omega_{F2})^3}
 \int_0^\infty dp_1 \int_0^\infty dp_2\, (p_1 p_2)^{\frac{N-3}{2}}e^{-\frac{N}{4J^2}(p_1^2+p_2^2)+\frac{iNE}{2J^2}(p_1-p_2)} \\
&\times\int_{-\infty}^\infty dr_1 \int_{-\infty}^\infty dr_2\, (r_1 r_2)^{N-2}\,e^{-\frac{N}{2J^2}(r_1^2+r_2^2)-\frac{iNE}{J^2}(r_1+r_2)}  \\
& \times (r_1-r_2) (p_1+p_2) (r_1+ p_1)(r_2+ p_1)(r_1- p_2)(r_2- p_2) \\
& \times e^{\frac{(r_1+r_2)(\omega_{F1}+\omega_{F2})}{2J^2}}\,  e^{-\frac{(p_1-p_2)(\omega_{B1}+\omega_{B2})}{4J^2}}\, K_0 \left( \frac{(\omega_{B1}-\omega_{B2})(p_1+p_2)}{4J^2} \right)  \\
&\times \left[\frac{(r_1-r_2)(\omega_{F1}-\omega_{F2})}{2J^2}\cosh\left(\frac{(r_1-r_2)(\omega_{F1}-\omega_{F2})}{2J^2}\right) \right. \\
&\phantom{\times\bigg[}\left. -\sinh\left(\frac{(r_1-r_2)(\omega_{F1}-\omega_{F2})}{2J^2}\right) \right].
\end{split}
\end{equation}
The superscript $+-$ is to remind us that the expression corresponds to the choice $\omega_{B1}>0$ and $\omega_{B2}<0$. The expression for equal signs can be obtained from the above  by replacing $p_1 \to +\sgn (\omega_{B1}) p_1$, $p_2 \to -\sgn (\omega_{B1}) p_2$, $p_1+p_2 \to |p_1 -p_2|$ and $K_0 \to I_0$.

So far our manipulations were exact and did not use any approximation. As was explained in the introduction we are mainly interested in extracting the ``bulk'' large-$N$ asymptotic of these correlation functions.  The most natural way to proceed from here is by performing a saddle-point analysis. We believe with due effort such analysis can be done with full mathematical rigor, see e.g. a recent paper \cite{Shcherbina2014}, but we do not attempt it here concentrating on explaining the gross structures of the saddle-point analysis which yield the correct results.

For the case of different signs the saddle points of the integrand are given by
\begin{equation}
\begin{split}
 p_1^{SP} &= \frac{iE + \sqrt{4J^2-E^2}}{2},\ p_2^{SP} = \frac{-iE + \sqrt{4J^2-E^2}}{2}, \\
 q^{SP}&=r_{1,2}^{SP}= \frac{-iE \pm \sqrt{4J^2-E^2}}{2}.
\end{split}
\end{equation}
For $p_1$ and $p_2$ only solutions with positive real parts are contributing to the asymptotics. There is no such restriction for $q$ or $r_1$ and $r_2$, respectively, and we have two saddle points contributing in each of these variables. Hence for $\mathcal{C}^{(\text{bulk},+-)}_{1,2}$ the final expression  is given by the sum of two different saddle-point contributions. For $\mathcal{C}^{(\text{bulk},+-)}_{2,2}$ there are in principle four different contributions. However, the contributions from the saddle points satisfying $r_1^{SP}=r_2^{SP}$ are actually negligible due to the factor $r_1-r_2$ in the integrand. Moreover the integrand is invariant under exchanging $r_1$ and $r_2$, and hence the two remaining contributions are identical. It therefore suffices to choose for $r_1^{SP}$ the solution with positive real part and for $r_2^{SP}$ the one with negative real part.
 One may further notice that the integrand itself vanishes when evaluated at the saddle points due to the factors $(q-p_1)(q+p_2)$ and $(r_1+ p_1)(r_2+ p_1)(r_1- p_2)(r_2- p_2)$ . This fact makes it necessary to expand the integrand to a higher order around the saddle points. The corresponding calculation is rather tedious, but  managable. We refrain from presenting it here and refer the interested reader to \cite{thesis} for technical detail.
The outcome of the analysis are precisely the formulae given in Eqs. \eqref{C1diff} and \eqref{C2diff}.

The case of same signs looks quite different. Here the saddle points are given by
\begin{equation}
 p_1^{SP}=p_2^{SP}=\frac{i s E+\sqrt{4J^2-E^2}}{2}, \quad q^{SP}=-r_{1,2}^{SP}=\frac{iE\pm\sqrt{4J^2-E^2}}{2},
\end{equation}
where $s \equiv \sgn\omega_{B1}=\sgn\omega_{B2}$. Again we must choose $p_1^{SP}$ and $p_2^{SP}$ to have a positive real part, so that two contributions arise for $\mathcal{C}^{(\text{bulk})}_{1,2}$ and four for $\mathcal{C}^{(\text{bulk})}_{2,2}$. However, the term $(q-s p_1)(q-s p_2)$ is only nonvanishing if we choose $q^{SP}=-sp_1^{SP}$,  contributions for all other choices becoming subdominant. For $\mathcal{C}^{(\text{bulk})}_{2,2}$ the same arguments as before suggest to choose for $r_1^{SP}$ the solution with positive real part and for $r_2^{SP}$  with negative real part neglecting the other three contributions. While the integrand still vanishes at he saddle points due to the factor $|p_1-p_2|$ and for $\mathcal{C}^{(\text{bulk})}_{2,2}$ due to the factors $(r_1+sp_1)(r_2+sp_1)(r_1+sp_2)(r_2+sp_2)$, the saddle point analysis is now much simpler than in the previous case. Indeed, when extracting the leading-order contribution one has to replace $p_1=p_1^{SP}+\xi_1$ (with $\xi_1$ parametrizing  the integration around the relevant saddle point) and similarly for the other variables, and then expand the $N$-independent part of the integrand to zero-th order in $\xi_1$ etc. (apart from the factors which come naturally in first order like $|p_1-p_2|=|\xi_1-\xi_2|$). It is then readily seen that the corresponding  integrals yield  a nonvanishing contribution rather straightforwardly without need to expand the integrand to higher orders like it was necessary in the previous case of opposite signs. The results of such saddle-point analysis  is then much simpler and is given in Eqs. \eqref{C1same} and \eqref{C2same}.

\subsection{Distribution of $K_{ab}$ via Eq.(\ref{C4special}).}

For the correlation function (\ref{C4special}) associated with the distribution of an individual off-diagonal $K$-matrix element we consider for simplicity only the perfect coupling case $E=0$ and $\gamma_a=\gamma_b=1$, see Eq. \eqref{RKab}). For evaluating the ensemble average we first tried to follow the same method as described in the previous section. In this way we started with writing $\det(H^2+\frac{x^2}{N^2})^{1/2}=\det(H+\frac{ix}{N})^{1/2}\det(H-\frac{ix}{N})^{1/2}$ and $|\det H|=(\det H)^2 /|\det H| = \lim_{\epsilon \to 0} (\det H)^2 \det(H+\frac{i\epsilon}{N})^{-1/2}\det(H-\frac{i\epsilon}{N})^{-1/2}$ and then replaced the square roots of characteristic polynomials in the denominator by four Gaussian integrals over real commuting vectors and those in the numerator by Gaussian integrals over two  vectors with anticommuting components.
 The ensemble averaging then yields a $4 \times 4$ $Q_B$-matrix, but we found no efficient ways of evaluating the ensuing group integral over the diagonalizing matrices. We also attempted a direct saddle-point analysis for large $N$ along the same lines as before, and found it to become very tedious as not only the zero-th and first, but also the second order of the integrand expansion in fluctuations around the relevant saddle points turned out to be vanishing at the saddle points. Expanding to an even higher order with the group integrals still present did not seem to us as a viable option.

 Confronted with those difficulties we followed a different method (inspired by the insights from \cite{SFPB}) which avoids introducing anticommuting variables altogether. We demonstrate it first for the correlation function $\mathcal{C}^{(\text{bulk})}_{1,2}$. For brevity we will consider only the simplest case $E=0$ where such object can be written as
 \begin{equation}
 \mathcal{C}^{(\text{bulk})}_{1,2}(\omega_{F};\omega_{B1},\omega_{B2})= \left\langle \frac{\det(\frac{i\omega_{F}}{N}-H)}{\det^{1/2}(H^2-\frac{\omega_{B1}\omega_{B2}}{N^2}-iH \frac{\omega_{B1}+\omega_{B2}}{N})} \right\rangle_{GOE, N \to \infty}.
 \end{equation}
We start with representing only the denominator by a Gaussian integral over a real $N$-component vector ${\bf S}$ and hence get
\begin{equation}
 \mathcal{C}^{(\text{bulk})}_{1,2}= \int d{\bf S}\, e^{\frac{\omega_{B1} \omega_{B2}}{2N^2}{\bf S}^2 } \Phi({\bf S},\omega_F,\omega_{B1}+\omega_{B2}),
\end{equation}
where
\begin{equation}\label{phi}
 \Phi({\bf S},\omega_F,\omega_{B1}+\omega_{B2})=\left\langle \frac{\det(\frac{i\omega_{F}}{N}-H)}{(2\pi)^{N/2}} \exp\left[-\frac{1}{2} {\bf S}^T \left(H^2-iH\frac{\omega_{B1}+\omega_{B2}}{N}\right) {\bf S} \right] \right\rangle_{GOE, N\to \infty}.
\end{equation}
 Note that the above integral is well-defined only for $\omega_{B1}$ and $\omega_{B2}$ having different signs, otherwise  the term $\omega_{B1} \omega_{B2}/N^2>0$ would render the integral divergent.

Let us further assume that $\omega_{B1}=-\omega_{B2} \equiv \omega_B$, such that the linear term $-iH\frac{\omega_{B1}+\omega_{B2}}{N}$ vanishes. Such assumption is not necessary to make the method functional but helps to simplify the presentation considerably. Next we parametrize the vector ${\bf S}$ of integration variables as ${\bf S}=|{\bf S}| O e_1$, where $e_1=[1,0,\dots,0]$ is an $N$-dimensional unit vector and $O$ is an orthogonal matrix: $O^{-1}=O^{T}$. Since both the determinant factor and the GOE  probability density $\mathcal{P}(H)$ in (\ref{phi}) are invariant under orthogonal transformations $H\to O^{-1}HO$ the matrices $O, O^T$ can be omitted. The term $e_1^T H^2 e_1$ then suggests that it is advantageous to decompose $H$ as
\begin{equation} \label{c1decomp}
 H=\begin{bmatrix}
  H_{11} & h^T \\
  h & H_{N-1}
 \end{bmatrix},
\end{equation}
where $h$ is a real $N-1$-component vector, $H_{N-1}$ is the $(N-1) \times (N-1)$ subblock of $H$ and $H_{11}$ is the first element of $H$. With such a decomposition one is able to integrate out the variable $H_{11}$ as well as the vector $h$, which leads to
\begin{equation}
 \mathcal{C}^{(\text{bulk})}_{1,2} \propto \int d{\bf S}
  \frac{ \frac{i\omega_F}{N}I_1  - \frac{1}{|{\bf S}|^2+\frac{N}{J^2}}I_2}{(|{\bf S}|^2+\frac{N}{J^2})^{\frac{N-1}{2}}(|{\bf S}|^2+\frac{N}{2J^2})^{1/2}} \exp \left[-\frac{\omega_B^2}{2N^2}|{\bf S}|^2 \right],
\end{equation}
where we have introduced the short-hand notations $I_1=\langle \det(\tfrac{i\omega_F}{N}-H_{N-1}) \rangle_{N-1}$ and $I_2=\left\langle \det(\tfrac{i\omega_F}{N}-H_{N-1})\ \tr(\tfrac{i\omega_F}{N}-H_{N-1})^{-1} \right\rangle_{N-1}$
where the ensemble average should be performed over the $(N-1)\times (N-1)$ GOE matrix $H_{N-1}$. Moreover, it actually suffices to know only $I_1$ since  $I_2=-iN \frac{dI_1}{d\omega_F}$. As is well-known $I_1$ is proportional to the Hermite polynomial: $I_1 \propto H_{N-1}(i\omega_F/(\sqrt{N}J))$, so that asymptotically we have $I_1 \propto e^{\omega_F/J}+(-1)^N e^{-\omega_F/J}$. It remains to perform the ${\bf S}$-integration for which it is advantageous to introduce rescaled polar coordinates, such that $|{\bf S}|^2=N^2 R$. The problem then reduces to performing the single integral
\begin{equation}
 \mathcal{C}^{(\text{bulk})}_{1,2} \propto \int_0^\infty \frac{dR}{R}\,
  \frac{\omega_F I_1 + \frac{1}{R(1+\frac{1}{N J^2R})}\frac{dI_1}{d\omega_F}}{(1+\frac{1}{NJ^2R})^{\frac{N-1}{2}}(1+\frac{1}{2NJ^2R})^{1/2}} \exp \left[-\frac{\omega_B^2}{2}R \right].
\end{equation}
For large $N\gg 1$ it is easy to verify that the leading contribution to the integral can be written as
\begin{align}
 \mathcal{C}^{(\text{bulk})}_{1,2} &\propto \int_0^\infty \frac{dR}{R}\,
  \left(\omega_F I_1  + \frac{1}{R}\frac{dI_1}{d\omega_F} \right) \exp \left[-\frac{\omega_B^2}{2}R -\frac{1}{2J^2 R}\right] \nonumber \\
  &\propto (e^{\frac{\omega_F}{J}}+(-1)^N e^{-\frac{\omega_F}{J}}) \omega_F K_0(\tfrac{|\omega_B|}{J})+(e^{\frac{\omega_F}{J}}-(-1)^N e^{-\frac{\omega_F}{J}}) |\omega_B| K_1(\tfrac{|\omega_B|}{J}),
\end{align}
which indeed coincides with the earlier derived expression for $\mathcal{C}^{(\text{bulk})}_{1,2}(\omega_F; \omega_B,-\omega_B)$ from Eq. \eqref{C1diff}.

Now we follow the same route for evaluation of the correlation function (\ref{C4special}).   We will only outline the key steps and differences from the previous case, but refrain from presenting intermediate results relegating them to \cite{thesis}. One starts with replacing only the square roots of the characteristic polynomials in the denominator by Gaussian integrals, which leads us to
\begin{equation} \label{S1S2}
 R(x)= \lim_{\epsilon \to 0} \frac{1}{(2\pi)^N} \int\! d{\bf S}_1\! \int\! d{\bf S}_2\, e^{-\frac{1}{2N^2}(x^2 {\bf S}_1^T {\bf S}_1+\epsilon^2 {\bf S}_2^T {\bf S}_2)}\, \Psi({\bf S}_1,{\bf S}_2),
\end{equation}
where ${\bf S}_1$ and ${\bf S}_2$ are two real $N$-component vectors, and
\begin{equation}\label{phi2}
 \Psi({\bf S}_1,{\bf S}_2)= \left\langle \det H^2\, e^{-\frac{1}{2}\tr H^2 Q} \right\rangle, \quad Q={\bf S}_1 \otimes {\bf S}_1^T + {\bf S}_2 \otimes {\bf S}_2^T.
\end{equation}
In contrast to a single vector ${\bf S}$ in the previous case we now have to deal with two real vectors ${\bf S}_1$ and ${\bf S}_2$, which we can conveniently combine into the matrix $Q$. Such a rank-two $N\times N$ matrix has two nonzero eigenvalues which we call $q_1$ and $q_2$, all other $N-2$ eigenvalues being identically zero. Being real symmetric $Q$ can be diagonalised by an orthogonal transformation: $Q=O \diag(q_1,q_2,0,\dots,0) O^T$ and the orthogonal matrices can be omitted from the integrand by the same invariance reasons as before. Owing to this structure we can conveniently decompose $H$ into its upper left $2 \times 2$ block, its lower right $(N-2) \times (N-2)$ block $H_{N-2}$ and the two ensuing off-diagonal blocks.  It is easy to integrate out all variables apart from those entering $H_{N-2}$ and get, with a slight abuse of notations:
\begin{equation*}
\Psi({\bf S}_1,{\bf S}_2)\to \Psi(q_1,q_2)=\frac{(2\pi)^{3/2}}{\sqrt{a_1a_2a_3}}\left(\frac{4\pi^2}{\sqrt{c_1c_2}}\right)^{N/2}\left\langle
\det{H^2_{N-2}}\left\{\frac{1}{a_1a_2}+\frac{3}{(a_1+a_2)^2}\right.\right.
\end{equation*}
\begin{equation}\label{psi}
\left.+\left(\frac{1}{a_2c_1^2}+\frac{1}{a_1c_2^2}\right)\left[2\mbox{tr}A^2+(\mbox{tr}A)^2\right]+\frac{2}{c_1c_2(a_1+a_2)}
\left[3\mbox{tr}A^2-(\mbox{tr}A)^2\right]
\right.
\end{equation}
\begin{equation*}
\left.\left.+\frac{1}{c_1^2c_2^2}\left((\mbox{tr}A)^4-8\mbox{tr}A\,\mbox{tr}A^3+7(\mbox{tr}A^2)^2+2(\mbox{tr}A)^2\,
\mbox{tr}A^2
-2\mbox{tr}A^4\right)\right\}\right\rangle_{GOE,N-2}
\end{equation*}
where we used the notations
\[
A=H^{-1}_{N-2}, \quad a_{1,2}=q_{1,2}+\frac{N}{J^2}, \quad c_{1,2}=q_{1,2}+2\frac{N}{J^2}
\]

The result then reduces to performing ensemble averages over expressions $\det H_{N-2}^2$ multiplied with various powers of traces of the inverse matrices $H_{N-2}^{-k}$ for a few instances of positive integers $k$.  One may notice that all the required averages can be represented as derivatives of the correlation function of two GOE characteristic polynomials, using e.g. the identities
\[
\det{H_{N-2}^2}\left((\mbox{tr}H_{N-2}^{-1})^2-\mbox{tr}H^{-2}_{N-2}\right)= \lim_{\xi_1,\xi_2 \to 0}\frac{\partial^2}{\partial \xi_1^2}\left[ \det(H_{N-2}-\xi_1)\det(H_{N-2}-\xi_2) \right]
\]
\[
\det{H_{N-2}^2}\left((\mbox{tr}H_{N-2}^{-1})^2\right)= \lim_{\xi_1,\xi_2 \to 0}\frac{\partial^2}{\partial \xi_1\partial x_2}\left[ \det(H_{N-2}-\xi_1)\det(H_{N-2}-\xi_2) \right]
\]
and similarly for the higher powers. As a result for the object featuring in (\ref{psi}) we have:
\begin{equation}
 \Psi(q_1,q_2) = \lim_{\xi_1,\xi_2 \to 0} \mathcal{D}_{\xi_1,\xi_2}(q_1,q_2) [\langle \det(H_{N-2}-\xi_1)\det(H_{N-2}-\xi_2) \rangle_{GOE,N-2},
\end{equation}
where the differential operator $\mathcal{D}_{\xi_1,\xi_2}(q_1,q_2)$ is explicitly given by
\begin{equation}
\begin{split} \label{diffop}
\mathcal{D}_{\xi_1,\xi_2}(q_1,q_2)= &\frac{(2\pi)^{3/2}}{\sqrt{a_1 a_2 (a_1+a_2)}} \left(\frac{4\pi^2}{c_1 c_2}\right)^{\frac{N}{2}-1}
  \Bigg\{ \left(\frac{1}{a_2 c_1^2}+\frac{1}{a_1 c_2^2}\right) \left(3 \frac{\partial^2}{\partial \xi_1 \partial \xi_2} -2 \frac{\partial^2}{\partial \xi_1^2} \right)   \\
   & + \frac{2}{c_1 c_2 (a_1+a_2)}  \left(2 \frac{\partial^2}{\partial \xi_1 \partial \xi_2} -3 \frac{\partial^2}{\partial \xi_1^2}\right)
+ \left(\frac{1}{a_1 a_2} + \frac{3}{(a_1+a_2)^2}\right) \\
   &+ \frac{1}{3c_1^2 c_2^2} \left(\frac{\partial^4}{\partial \xi_1^4}+18\frac{\partial^4}{\partial \xi_1^2 \partial \xi_2^2} -16 \frac{\partial^4}{\partial \xi_1^3 \partial \xi_2} \right) \Bigg \}.
\end{split}
\end{equation}

 The ensemble average of the product of two GOE characteristic polynomials is known and for large $N$ is given asymptotically by (see e.g. \cite{Koesters08})
 \begin{equation} \label{Dlimit}
 \langle \det(H_{N-2}-\xi_1)\det(H_{N-2}-\xi_2) \rangle_{GOE} \propto
 \frac{\sinh\left(\frac{\xi_1-\xi_2}{J}\right)-\frac{\xi_1-\xi_2}{J}\cosh\left(\frac{\xi_1-\xi_2}{J}\right)}{(\xi_1-\xi_2)^3}.
\end{equation}
Using this result, and taking the necessary derivatives and the limits $\xi_1, \xi_2 \to 0$, we finally get an explicit expression for $\Psi(q_1,q_2)$.

The last step is to perform the integrals over ${\bf S}_1$ and ${\bf S}_2$, see Eq. \eqref{S1S2}. In the previous case we could reduce integration over ${\bf S}_1$ to a single integration in polar coordinates. Similarly we can now exploit the invariance of the integrand and exploit the identity \eqref{IT}. In this way we can restrict the integration to the manifold of positive definite real symmetric $2 \times 2$ matrices with eigenvalues $q_1$ and $q_2$.  Extracting the leading large-$N$-asymptotics is then a straightforward exercise and we finally end up with the integral representation
\begin{equation}
\begin{split} \label{RxInt}
 R(x) \propto &\int_0^\infty dq_1 \int_0^\infty dq_2\, \frac{|q_1-q_2|}{q_1 q_2 \sqrt{q_1+q_2}}\, I_0\left[\frac{(x^2-\epsilon^2)(q_1-q_2)}{4J^2} \right] \\
 &\times \exp \left[-\frac{1}{2}\left(\frac{1}{q_1}+\frac{1}{q_2}+\frac{(q_1+q_2)(x^2+\epsilon^2)}{2J^2}\right)\right] \\
  &\times\Bigg\{\frac{(1+q_1)(1+q_2)}{q_1^2 q_2^2}+ \frac{3}{(q_1+q_2)^2} + \frac{2}{q_1 q_2 (q_1+q_2)} \Bigg\}.
 \end{split}
\end{equation}
Note that here the limit $\epsilon \to 0$ is implied, which can now trivially be performed. It turns out that this rather complicated-looking integral is actually proportional to
\begin{equation} \label{Rx}
 R(x) \propto \frac{|x|}{J} K_0(|x|/J) + \int_{|x|/J}^\infty dy\, K_0(y).
\end{equation}
A way to verify this claim is to differentiate both equations (assuming for definiteness $x>0$, $J=1$) with respect to $x$. The derivative of Eq. \eqref{Rx} is $x K_1(x)$, and the derivative of \eqref{RxInt} is $x$ times a certain two-fold integral which with some efforts can be shown to be proportional to $K_1(x)$. The details of this calculation are relegated to \cite{thesis}.

\section{Conclusions and open problems}

In this paper we have started the program of systematic evaluation of correlation functions (\ref{1})
involving half-integer powers of the characteristic polynomials of $N\times N$ GOE matrices. Motivated
 by diverse applications outlined in the introductory section we mainly concentrated on extracting the asymptotic behaviour of several objects of that type as $N\to \infty$. Our calculations were based on variants of the supersymmetry method or related techniques. The method in a nutshell amounts to replacing the initial average involving the product of $K$ characteristic polynomials divided by $L$ square roots of characteristic polynomials of $N\times N$ GOE matrices $H$  with an average over the sets of $K\times K$ matrices $Q_F$ and $L\times L$ matrices $Q_B>0$  with Gaussian weights augmented essentially with the factors $\det{Q_B}$ and $\det{Q_F}$ raised to powers of order $N$, see e.g. (\ref{c2withQ}). As we are eventually mostly interested in $K,L$ fixed but $N\to \infty$ this replacement is very helpful as it allows to employ saddle-point approximations. In this paper we managed to perform all steps of such a procedure successfully only for relatively small values of $K$ and $L$, but we hope that  the general case can eventually be treated along similar lines. One reason and guiding principle for a moderate optimism
 is as follows. An inspection of a somewhat simpler example of $\beta=2$ shows, see in particular \cite{FS03a}, that the success of  our method is deeply connected to the existence of the so-called {\it duality relations} for Gaussian ensembles, see \cite{Des2008} for a better understanding of such dualities. In particular, the Proposition 7 of the latter paper shows that one of such duality relations exists for general Gaussian $\beta$-ensembles with $\beta>0$ for an object involving the ensemble average of the product of the corresponding characteristic polynomials raised to the power $-\beta/2$. For the GOE with $\beta=1$ that object (see Proposition 2 in \cite{Des2008}) is exactly  the particular case of (\ref{1}) with $K=0$ and arbitrary integer $L$ which makes a contact to the present context; e.g. one can employ such a duality to reproduce the relation (\ref{fyokeat}) in an alternative way. A deeper understanding of connections between the supersymmetric approach and the duality relations for Gaussian ensembles will certainly be helpful in dealing efficiently with asymptotics of (\ref{1}) for arbitrary integer values $K$ and $L$. The problem of revealing possible Pfaffian-determinant structures behind (\ref{1}) for finite matrix size $N$ remains at the moment completely outstanding. It may well be that the methods of \cite{KG1,KG2} or relations to generalized hypergeometric functions noticed for some particular instances in \cite{ForrKeat} could be useful for clarifying that issue.

\section*{Acknowledgements}
 Y.V.F. and A.N. were supported by EPSRC grant EP/J002763/1 ``Insights into Disordered Landscapes via Random Matrix Theory and Statistical Mechanics''.
\vspace{\baselineskip}

\appendix
\section{Evaluation of the distribution for $K_{ab}$ using Brouwer's conjecture.} \label{app1}
We show that the matrix Cauchy-type probability density $\mathcal{P}(K) \propto \det[1+K^2]^{-(M+1)/2}$ leads to the same answer for the distribution of an off-diagonal matrix element as the Hamiltonian approach, given in Eq. \eqref{PKab}. Without loss of generality we may choose $M=2$ when we have explicitly
\begin{equation}
 \mathcal{P}(K) \propto [(1+K_{11}^2)(1+K_{22}^2)+2K_{12}^2 (1-K_{11}K_{22})+K_{12}^4]^{-3/2}.
\end{equation}
 In order to obtain the probability density for $K_{12}$ we need to integrate out the other two variables. We start with integrating out the variable $K_{22}$. The integrand is of the form $(a K_{22}^2+b K_{22}+c)^{-3/2}=\left[a(K_{22}+\frac{b}{2a^2})^2-\frac{b^2}{4a}+c\right]^{-3/2}$ with $a=1+K_{11}^2,\ b=-2K_{11}K_{12}^2,\ c=1+K_{11}^2+2K_{12}^2+K_{12}^4$. Now we change variables $\sqrt{\frac{a}{D}}(K_{22}+\frac{b}{2a^2}) \to K_{22}$ where we denoted $D=c-\frac{b^2}{4a}=\frac{(1+K_{11}^2+K_{12}^2)^2}{1+K_{11}^2}>0$.  The joint probability density of $K_{11}$ and $K_{12}$ is then given by
\begin{equation}
 \mathcal{P}(K_{11},K_{12}) \propto \frac{1}{\sqrt{a} D} \int_{-\infty}^{+\infty}  \frac{dK_{22}}{(1+K_{22}^2)^{3/2}} = \frac{2}{\sqrt{a}D} =  \frac{2\sqrt{1+K_{11}^2}}{(1+K_{11}^2+K_{12}^2)^2}.
\end{equation}
To integrate out $K_{11}$  we change variables $K_{11}=\frac{y}{a}\sqrt{\frac{1}{1-y^2/a^2}}$, with $a=\frac{K_{12}}{\sqrt{1+K_{12}^2}}$. As the integrand is even the integral transforms to
\begin{equation}
 \int_{-\infty}^{+\infty} dK_{11} \frac{\sqrt{1+K_{11}^2}}{(1+K_{11}^2+K_{12}^2)^2} \propto \frac{1}{K_{12}(1+K_{12}^2)^{3/2}} \int_0^a \frac{dy}{(1-y^2)^2}.
\end{equation}
The integration on the right-hand side can be easily performed as
\begin{equation}
\int_0^a \frac{dy}{(1-y^2)^2} = \frac{a}{1-a^2} - \int_0^a \frac{y^2}{(1-y^2)^2} dy = \frac{1}{2} \left(\frac{a}{1-a^2} +\int_0^a \frac{dy}{1-y^2} \right),
\end{equation}
with the last integral on the right yielding $\text{artanh}\,a$. In this way we arrive at the probability density for $K_{12}$ in the form
\begin{equation}
 \mathcal{P}(K_{12}) \propto \frac{1}{K_{12}(1+K_{12}^2)^{3/2}} \left(\frac{a(K_{12})}{1-a^2(K_{12})} + \text{artanh}\,a(K_{12}) \right).
\end{equation}
It can be finally brought to the form of Eq. \eqref{PKab} by reinserting $a(K_{12})=\frac{K_{12}}{\sqrt{1+K_{12}^2}}$ and employing the identity $\text{artanh} \left(\frac{x}{\sqrt{1+x^2}}\right)=\text{arsinh}\,x$.

\section{Consistency between Eq. \eqref{M=2b} and Brouwer's conjecture} \label{app2}
We show that the characteristic function of the probability density $\mathcal{P}(K)$ in the case $M=2$ given in Eq. \eqref{M=2b} is fully consistent with the claim that $\mathcal{P}(K) \propto \det[1+K^2]^{-3/2}$.
For the particular choice $\gamma_1 x_1 = -\gamma_2 x_2 \equiv \gamma x$ the expression Eq. \eqref{M=2b} is equivalent to Eq. \eqref{RKab} (for brevity we choose $\gamma=1$). Our task then amounts to demonstrating that
\begin{equation} \label{claim}
\int dK e^{\frac{i}{2}\Tr KX}\det[1+K^2]^{-3/2} \propto x K_0(x)+\int_x^\infty dy K_0(y),
\end{equation}
where $X$ can be chosen diagonal, $X=\diag(x,-x)$. Since $K$ is symmetric we can diagonalise it by an orthogonal transformation, $K=O\diag(k_1,k_2)O^T$. Choosing for $O$ the standard parametrization of a $2 \times 2$ orthogonal matrix, the left-hand side of Eq. \eqref{claim} then simplifies to
\begin{equation} \label{claimdiag}
\int_{-\infty}^\infty dk_1 \int_{-\infty}^\infty dk_2 \frac{|k_1-k_2|}{(1+k_1^2)^{3/2}(1+k_2^2)^{3/2}} \int_0^{2\pi} d\phi\, e^{\frac{i}{2}x(k_1-k_2)\cos(2\phi)}.
\end{equation}
The integral over the angle yields a Bessel function, and can also be rewritten in the form $ \int_0^{2\pi} d\phi\, e^{\frac{i}{2}x(k_1-k_2)\sin(2\phi)}$. Now note that $\frac{1}{2}(k_1-k_2)\sin(2\phi) \equiv -K_{12}$, which allows to  present Eq. \eqref{claimdiag} in the form
\begin{equation}
  \int dK e^{-ixK_{12}} \det[1+K^2]^{-3/2}.
\end{equation}
This is precisely the Fourier transform of $\mathcal{P}(K_{12})$, which  due to \ref{app1} is proportional to $ x K_0(x)+\int_x^\infty dy K_0(y)$. This shows the validity of the claim \eqref{claim}.
\vspace{\baselineskip}

\end{document}